\documentclass[preprint,aps,showpacs,floats,letterpaper,floatfix,groupedaddress,eqsecnum]{revtex4}
\usepackage{amssymb,amsmath}
\usepackage[dvips]{graphicx}
\usepackage{dcolumn,epsfig}

\newcommand{\beq}{\begin{equation}}
\newcommand{\eeq}{\end{equation}}
\newcommand{\bes}{\begin{subequations}}
\newcommand{\ees}{\end{subequations}}
\newcommand{\bea}{\begin{eqnarray}}
\newcommand{\eea}{\end{eqnarray}}
\newcommand{\ba}{\begin{array}}
\newcommand{\ea}{\end{array}}
\newcommand{\beqn}{\begin{eqnarray*}}
\newcommand{\eeqn}{\end{eqnarray*}}

\newcommand{\f}[2]{\frac{#1}{#2}}

\newcommand{\pa}{\partial}

\def\nn{\nonumber}

\newlength{\sizeonefig}
\newlength{\sizetwofig}
\begin{document}

\title{Decohering d-dimensional quantum resistance}

\author{Dibyendu Roy, N. Kumar} 
\email{dibyendu@rri.res.in, nkumar@rri.res.in}
\affiliation{Raman Research Institute, Bangalore 560080, India}

\begin{abstract}
The Landauer scattering approach to  4-probe resistance is revisited for the case of a d-dimensional disordered resistor $in~ the~ presence~ of~ decoherence$. Our treatment is based on an invariant-embedding equation for the evolution of the coherent reflection amplitude coefficient in the length  of a 1-dimensional disordered conductor, where decoherence is introduced $at~ par$ with the disorder through an outcoupling, or  stochastic absorption, of the wave amplitude into  side (transverse) channels, and its subsequent incoherent re-injection into the conductor. This is essentially in the spirit of B{\"u}ttiker's reservoir-induced decoherence. The resulting evolution equation for the probability density of the 4-probe resistance in the presence of decoherence is then generalised from the 1-dimensional to the d-dimensional case following an anisotropic Migdal-Kadanoff-type procedure and analysed. The anisotropy, namely that the disorder evolves in one arbitrarily chosen direction only, is the main approximation here that makes the analytical treatment possible. A qualitatively new result is that arbitrarily small decoherence reduces the localisation-delocalisation transition to a crossover making resistance moments of all orders finite. 

\end{abstract}

\vspace{0.5cm}
\date{\today}

\maketitle

\section{Introduction}
 
Electron localisation \cite{Anderson58, TVR85}, strong as well as weak, and the associated metal-insulator transition and  conductance fluctuations \cite{Kumar04} are due essentially to the time-persistent interference of the complex wave amplitudes that result from multiple elastic scattering on randomly distributed defects in the  conductor with quenched potential disorder. Similar interference effects also manifest in mesoscopic systems as various phase-sensitive phenomena, e.g. the well known persistent ring currents and the Aharonov-Bohm oscillations. Clearly, these one-electron phase-sensitive phenomena can get suppressed by decoherence. Microscopically \cite{Lerner04}, decoherence can arise from  incoherent processes involving, e.g., the inelastic electron-phonon or the electron-electron scattering, as also from  an entanglement with the environmental degrees of freedom that remain undetected or unmeasured. (While coherent inelastic scattering is in principle possible, as indeed in the case of  neutron scattering, it is not relevant to the case of coherent multiple scattering of electrons in a disordered conductor). The question now is how to incorporate decoherence phenomenologically in an analytical treatment of the otherwise Hamiltonian system such as the system of non-interacting electrons moving in a lattice, or a continuum  with random elastic scatterers, e.g., the Anderson model system for metal-insulator transition in random lattices \cite{Anderson58}. Decoherence has often been included theoretically and probed experimentally through a phase breaking, or dephasing cut-off length scale introduced on physical grounds \cite{TVR85, Bergmann84}. It is clearly desirable, however, to have a phenomenology  for introducing the degree of decoherence  in the analytical treatment of elastic scattering in a disordered conductor. A highly successful and widely used approach to decoherence was pioneered by B{\"u}ttiker \cite{Buttiker85, Buttiker86, Pilgram06, Forster07}  through the idea of reservoir-induced decoherence. The latter could be introduced naturally in the scattering approach of Landauer \cite{Landauer70} to quantum transport, e.g., the 4-probe resistance. For the reservoir-induced decoherence, one inserts a scattering matrix with appropriately chosen side (transverse) channels, and thereby outcouple  a partial wave amplitude into an electron reservoir. The amplitude re-emitted from the reservoir is then re-injected back into the conductor, adding necessarily incoherently to the transmitted amplitude along the conductor (the longitudinal channel) that carries the transport current. The chemical potential of the reservoir is, of course, tuned so as to make the net current in the side channel vanish on the average. (This is clearly analogous to the ``potentiometric'' probe of Engquist and Anderson \cite{Engquist81}). The net result is the introduction of decoherence, or partial coherence, that can be readily parametrised. It describes, in particular, the quantum-to-classical crossover of a series combination of conductors \cite{Buttiker86} with increasing strength of the coupling to the intervening reservoirs. While used extensively in the context of mesoscopic (zero-dimensional) systems \cite{Imry97}, the reservoir-induced decoherence has also been invoked by many workers for treating partial coherence in quantum transport on tight-binding lattices -- without disorder \cite{Datta89, Datta91, DibAbhi07}, and with weak disorder \cite{Pastawski90, Maschke91, Maschke94}, as also in a disordered continuum \cite{RoyKumar07}. These studies are, however, confined to 1-dimensional conductors. 

In this work, we have considered the case of a d-dimensional conductor for $d\ge1$ in the presence of both the quenched disorder and decoherence. Our analytical treatment is based on the invariant-embedding approach developed earlier for a 1-dimensional conductor with quenched disorder \cite{Kumar85, Heinrichs86, Rammal87}, and its subsequent generalisation to higher dimensions for anisotropic disorder using the Migdal-Kadanoff technique \cite{Kumar86, Shapiro86}. Here, first the elastic scatterers (resistances) are combined in series quantum-mechanically along  an arbitrarily chosen direction and then classical Ohm's law is used to combine these resistances in parallel along the transverse directions. This is followed by a scaling transformation  with  an infinitesimal increase in scale at each step. The resulting `transverse' mixing up of disorder through the evolution equation is known to give a qualitatively correct description of the weak scattering regime in the absence of decoherence \cite{Shapiro86}, despite the assumption of anisotropic disorder, which is an approximation. In our approach, decoherence and elastic scattering (quenched disorder) are treated formally $at~ par$ through a proper insertion of the scattering ($S$-) matrices, i.e., transverse channels distributed over the conductor. Specifically, a side-channel is to be viewed as causing a stochastic absorption -- a coherent process by itself. The incoherent re-injection with zero net side-current is, however, effectively realised through the use of the Landauer expression $|R(L)|^2/(1-|R(L)|^2)$ for the 4-probe resistance, but with $|R(L)|^2$ now calculated as the coherent-only reflection coefficient. A physically robust argument is presented for the self-consistency of this procedure. The main results derived are, (a) elimination of the metal-insulator transition (the unstable fixed point) for an arbitrarily small strength of decoherence. This is indeed expected on physical grounds inasmuch as metal-insulator transition with increasing static disorder is essentially due to the coherent-back scattering \cite{TVR85} (where the back scattered amplitudes traversing the time-reversed paths add up in phase), while decoherence suppresses this phase coherent effect; (b) suppression of the 4-probe resistance fluctuations with increasing decoherence strength making all the resistance moments finite; and (c) a correction to conductivity due to decoherence in the metallic limit that mimics the conventional phase cut-off length scale.

\section{Model and invariant-embedding: 1-dimensional case}
\label{sec:model}

\begin{figure}[t]
\begin{center}
\includegraphics[width=15.0cm]{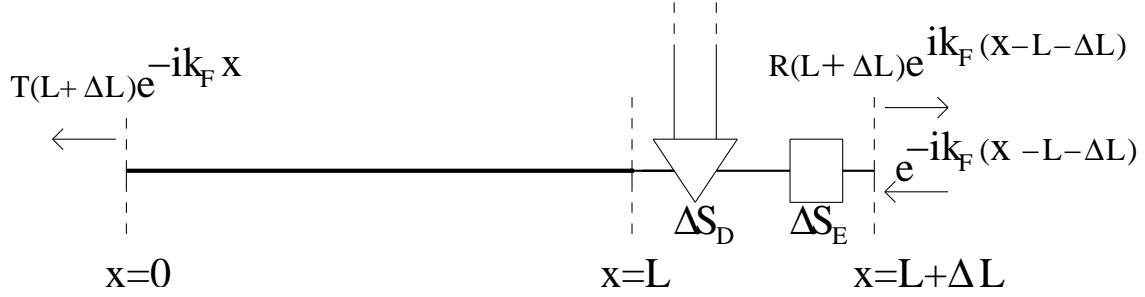}
\end{center}
\caption{Shows disordered sample of length L embedded invariantly in a supersample of length $L+\Delta L$. Shown also are the elementary  matrices for the elastic $(\Delta S_E)$ and the decohering $(\Delta S_D)$ scatterings in $\Delta L$, with the incident, the transmitted, and the reflected waves at Fermi wavevector $k_F$.}
\label{cartoon}
\end{figure}

Consider a model Hamiltonian $H$ for the system of non-interacting electrons in a 1-dimensional disordered conductor of length $L$: 
\bea
H=-\f{\hbar^2}{2m}\f{\pa^2}{\pa x^2}+V(x), 
\eea
where $V(x),~0<x<L$ is a spatially random potential (quenched disorder) assumed to be delta-correlated Gaussian as 
\bea
\langle V(x) V(x') \rangle={V_0}^2\delta(x-x')~. \nn 
\eea
Let an electron wave of unit amplitude be incident at  Fermi energy $(E_F=\hbar^2 k_F^2/2m)$ on the sample from  right, and let $R(L)$ and $T(L)$, respectively, be the reflection and the transmission amplitude coefficients. Next, let the sample of length $L$ be embedded invariantly in a supersample of length $L+\Delta L$ (Fig.~\ref{cartoon}). It is readily seen that the elastic scattering from the random potential in the interval $\Delta L$ with $k_F \Delta L << 1$ can be viewed as due to a delta-function potential of strength $V(L)\Delta L$, the corresponding scattering matrix being $\Delta S_E$
\begin{eqnarray}
\Delta S_E = \left(
\begin{array}{ll}
~~~\f{2mV\Delta L}{2 i\hbar^2 k_F} &~ 1+\f{2mV\Delta L}{2 i\hbar^2 k_F}\\
1+\f{2mV\Delta L}{2 i\hbar^2 k_F} &~~~~\f{2mV\Delta L}{2 i\hbar^2 k_F}
\end{array}
\right)
\label{SmatE}
\end{eqnarray}
This gives an evolution equation for the  $S$-matrix in the sample length L. Specifically, we have for the amplitude reflection coefficient \cite{Kumar85, Heinrichs86, Rammal87}
\bea
\f{dR}{dL}&=&i\f{k_F}{2}\xi(L)(1+R(L))^2 + 2ik_FR(L)~,\\
{\rm with}~~\xi(L)&=&-\f{2mV(L)}{\hbar^2 k_F^2}~~{\rm and}~~\langle \xi(L)\xi(L') \rangle = \Lambda \delta (L-L'). \nn
\eea
We are now in a position to introduce decoherence $at~ par$ with the random elastic scattering within this approach. We recall the $4\times4$ $S-$matrix with the side channels as introduced by B{\"u}ttiker \cite{Buttiker86}:
\begin{eqnarray}
S= \left(
\begin{array}{llll}
 ~~~~~0 &~~~ \sqrt{1-\epsilon} &~~~~~ \sqrt{\epsilon}&~~~~~~~0 \\
\sqrt{1-\epsilon} & ~~~~~~~0 &~~~~~~~0&~~~~~\sqrt{\epsilon} \\
~~~ \sqrt{\epsilon} &~~~~~~~0&~~~~~~~0&~-\sqrt{1-\epsilon}\\ 
~~~~~0 &~~~~~ \sqrt{\epsilon} &~- \sqrt{1-\epsilon}&~~~~~~~0
\end{array}
\right)
\end{eqnarray}
Here the outcoupling through the side channels is parametrised by $\epsilon$, which must be of order $\Delta L$ in the present case. Accordingly, we use the $2 \times 2$ sub-matrix 
\begin{eqnarray}
\Delta S_D = \left(
\begin{array}{ll}
~~~~0 & \sqrt{1-\epsilon}\\
\sqrt{1-\epsilon} &~~~~~0
\end{array}
\right)
\label{SmatD}
\end{eqnarray}
for insertion into the interval $\Delta L$.  It describes the outcoupling into the side channels, i.e., the stochastic absorption, as also the coherent transmission directly through the interval $\Delta L$. (Its connection with the reservoir-induced decoherence will be clarified below later). Figure (\ref {cartoon}) is a schematic depicting the insertion of the elementary  $\Delta S_E$ and $\Delta S_D$ in the interval $\Delta L$. Clearly, for $k_F \Delta L << 1$, the exact spatial order and the locations of the two insertions within the interval $\Delta L$ are not relevant.
Combining these two elementary S-matrices ($\Delta S_E$ and $\Delta S_D$) for $\Delta L$ with the $S-$matrix $(S(L))$ for the sample of length $L$ in series, we can read off the emergent quantities $R(L)$ and $T(L)$:
\bea
R(L+\Delta L)&=&\Delta R + \f{{\Delta T}^2~e^{2ik_F\Delta L}~R(L)}{1-\Delta R~R(L)~e^{2ik_F\Delta L}}~,\label{II1}\\
{\rm with}~~~\Delta R&=&\f{2mV\Delta L}{2 i\hbar^2 k_F}~{\rm and}
~~{\Delta T}^2=1-\epsilon+\f{2mV\Delta L}{i\hbar^2 k_F}.\nn
\eea
In the limit $\Delta L \to 0$, we obtain the evolution equations for the amplitude reflection/transmission coefficients $R(L)$ and $T(L)$:
\bea
\f{dR_c}{dL}&=&i\f{k_F}{2}\xi(L)(1+R_c(L))^2 + 2ik_FR_c(L)-\eta R_c(L)~,\label{II2} \\
{\rm and}~~\f{dT_c}{dL}&=&i\f{k_F}{2}\xi(L)(1+R_c(L))T_c(L) + ik_FT_c(L)-\f{\eta}{2} T_c(L)~,\label{IT2} 
\eea

${\rm where}~~\eta=\epsilon/\Delta L$, $\Delta L \to 0$ parametrises decoherence. Here we have introduced the subscript $`c$' just to emphasize that the reflection/transmission amplitude coefficients in Eq.(\ref {II2}) are coherent. 
 
It seems in order at this stage to clarify how decoherence is realised in relation to the sample resistance by the insertion of the side channel through $\Delta S_D$. Clearly, the embedding Eq.(\ref{II2}) describes  evolution of the coherent reflection amplitude $R_c(L)$. (Similarly, $T_c(L)$ is the coherent transmission amplitude as depicted in Fig.(\ref {cartoon}). The embedding equation for $T_c(L)$, however, is not autonomous -- it involves $R_c(L)$). The outcoupling into the side channels corresponds to a stochastic absorption \cite{Summhammer, Arun02, Pradhan06} in the interval $\Delta L$.  This, however, has to be re-injected now incoherently back into the conductor. Inasmuch as this re-injected current necessarily flows down the chemical potential gradient, it contributes to the total transmitted current equal to (within constant of proportionality) $|T_c(L)|^2+|T_{in}(L)|^2\equiv |T_{tot}(L)|^2$, where the subscript `in' denotes incoherent. From the conservation of the total current flowing down the conductor, we must have $|T_c(L)|^2+|T_{in}(L)|^2=1-|R_c(L)|^2$. Now, recall that the Landauer resistance $(\rho^{(d,D)})$ formula $\rho^{(d,D)}=(1-|T_{tot}|^2)/|T_{tot}|^2$ holds for arbitrary $|T_{tot}|^2$ (coherent or incoherent both). Here the superscript $(d,D)$ denotes the dimensionality $d$ and the decoherence parameter $D$.   Thus, we have $\rho^{(d,D)}=|R_c|^2/(1-|R_c|^2)$ given entirely in term of $R_c(L)$ which is calculable from Eq.(\ref {II2}). Thus the 4-probe resistance $|R_c(L)|^2/(1-|R_c(L)|^2)$ incorporates self-consistently  the incoherent re-injection. Here we must re-emphasize that $|R_c(L)|^2$ is the coherent reflection coefficient given by and calculable from the embedding Eq.(\ref {II2}). 

Our next step is to obtain the `Fokker-Planck' equation for the probability density of the reflection coefficient $r(L)=|R_c(L)|^2$ from the stochastic differential Eq.(\ref {II2}) which serves as the Langevin equation here. Following the now familiar procedure \cite{Kumar85, Heinrichs86, Rammal87, Pradhan94}, we obtain  
\bea
\f{\pa P^{(1)}(r,l)}{\pa l}&=&\f{\pa}{\pa r}[r \f{\pa}{\pa r}(1-r)^2 P^{(1)}(r,l)]+D \f{\pa}{\pa r}[r P^{(1)}(r,l)]~,\label{II3}\\
{\rm with}~~l&=&\f{L}{l_0}~,~~~l_0=\f{2}{\Lambda k_F^2}~~{\rm and}~~D=2\eta l_0~.\nn
\eea
This is clearly a two-parameter ($l_0~ {\rm and}~ D$) evolution equation. The two independent parameters $l_0$ and $D$ are, of course, composed of the two basic independent parameters $\Lambda k_F^2$ (measure of disorder) and $\eta$ (measure of decoherence). Thus, e.g., $D$ may vary through $\eta$, while $l_0$ can remain constant.

Equation (\ref {II3}) in the limit of large length $L>>l_0$ gives a steady-state distribution $P_{\infty}(r)$ for the reflection coefficient $r$
\bea
P^{(1)}_{\infty}(r)=\f{|D|\exp(|D|)\exp(-\f{|D|}{1-r})}{(1-r)^2}~,~~~~ r \le 1.\label{dist}
\eea
Note that for $D=0$, the limiting distribution in Eq.(\ref {dist}) tends to
 the delta- function, $\delta(1-r)$, and not to zero. (This can be
readily seen by noting that the probability distribution is
normalized to unity for all $D$). This  means that the reflection
coefficient becomes unity with probability one, as it must for
an  infinitely long $1d$ disordered wire without decoherence (all states being  localized then, a well known result from Anderson localization in one dimension). The corresponding resistance moments are all finite for $D \ne 0$. In particular, the limiting value of the average  4-probe resistance in the presence of decoherence is 
\bea
\rho^{(1,D)}_{\infty}&=&\f{\pi \hbar}{e^2}\langle\f{r}{1-r}\rangle \nn \\
&=&\f{\pi \hbar}{e^2|D|}~.
\eea
With this preparation (Eq.(\ref {II3})) in hand, we now turn to the case of d-dimensions.

\section{Higher-Dimensional case} 
\label{sec:hi}

Changing over to the 4-probe resistance $\rho=r/(1-r)$ (measured in the unit of $\pi \hbar/e^2$) as the new variable with the associated probability density $P^{(1)}(\rho,l)$, Eq. (\ref {II3}) reduces to
\bea
\f{\pa P^{(1)}}{\pa l}=\rho(\rho+1)\f{\pa^2 P^{(1)}}{\pa \rho^2}+\{(2\rho+1)+D\rho(\rho+1)\}\f{\pa P^{(1)}}{\pa \rho}+D(2\rho+1)P^{(1)}~.\label{II4}
\eea
The corresponding nth resistance moment in 1 dimension is 
\bea
\rho^{(1,D)}_n=\int^{\infty}_{0}P^{(1)}(\rho,l){\rho}^n d\rho~.
\eea
Multiplying both sides of Eq.(\ref {II4}) by ${\rho}^n$ and integrating by parts on the RHS, we get the evolution equation for the 1-dimensional moment 
\bea
\f{\pa \rho^{(1,D)}_n}{\pa l}=n(n+1)\rho^{(1,D)}_n +n^2\rho^{(1,D)}_{n-1}-Dn\rho^{(1,D)}_n-Dn\rho^{(1,D)}_{n+1}~,\label{rmom1}
\eea
which is hierarchical in nature (i.e., the equation for $\rho^{(1,D)}_n$ involves $\rho^{(1,D)}_{n-1}$ and $\rho^{(1,D)}_{n+1}$). For $D=0$, however, the equation for  $\rho^{(1)}_n$ involves the lower-order moments only leading to a closure of the hierarchy. Thus, the presence of decoherence $(D\ne0)$ brings about a qualitative change in the structure of the coupled equations for the moments of different orders. For $D=0$, the 
solutions of Eq.(\ref {rmom1}) for the 1st and the 2nd moments are readily obtained as 
\bea
\rho^{(1,0)}_1&=&\f{1}{2}(e^{2l}-1)~, \nn\\
\rho^{(1,0)}_2&=&\f{2}{3}(2{\rho^{(1,0)}_1}^3+3{\rho^{(1,0)}_1}^2)~.\label{rmom2}
\eea
In writing the last equation above, we have eliminated the length $l$ in favour of an implicit relation between $\rho^{(1,0)}_2~{\rm and}~\rho^{(1,0)}_1$. We have verified by iteration of Eq. (\ref{rmom1}), that this relation remains valid for $\rho^{(1,D)}_2~{\rm and}~\rho^{(1,D)}_1$ to a good approximation for $D\ne 0$, and will be used as such. Substituting for $\rho^{(1,D)}_2$ in terms of $\rho^{(1,D)}_1$ in Eq. (\ref{rmom1}) for $n=1$, and integrating we obtain a relation between $l$ and $\rho^{(1,D)}_1$
\bea
l=\int^{\rho^{(1,D)}_1}_0\f{d\tilde\rho^{(1,D)}_1}{-\f{4}{3}D{{\tilde{\rho}}^{{(1,D)}^3}}_1-2D{\tilde\rho^{{(1,D)}^2}}_1+(2-D)\tilde\rho^{(1,D)}_{1}+1}~.\label{lrelation}
\eea 
(From now dummy integration variable will be distinguished by a tilde). Hereinafter, the superscript $D$ in $\rho^{(1,D)}_1$ will be dropped except when required for the sake of clarity.
Defining the associated moment generating function $\chi^{(1)}(x,l)$ and the cumulant generating function  $K^{(1)}(x,l)$ of $P^{(1)}(\rho,l)$ as
\bea
\chi^{(1)}(x,l)&\equiv&\int^{\infty}_{0}e^{-x\rho}P^{(1)}(\rho,l)d\rho~, \nn \\
K^{(1)}(x,l)&\equiv&\ln \chi^{(1)}(x,l) ~,\nn 
\eea
we  derive from Eq.(\ref {rmom1}) their evolution equations
\bea
\f{\pa \chi^{(1)}}{\pa l}&=&(x^2+Dx)\f{\pa^2\chi^{(1)}}{\pa x^2}+(2x-Dx-x^2)\f{\pa \chi^{(1)}}{\pa x}-x\chi^{(1)}~, \label{momgen1}\\
\f{\pa K^{(1)}}{\pa l}&=&(x^2+Dx)\f{\pa^2K^{(1)}}{\pa x^2}+(x^2+Dx)(\f{\pa K^{(1)}}{\pa x})^2+(2x-Dx-x^2)\f{\pa K^{(1)}}{\pa x}-x~.\label{cumgen1}
\eea
Now, we proceed to generalise the above equations to the case $d>1$ . For this we closely follow the Migdal-Kadanoff procedure as in Ref.\cite{Kumar86}, assuming the quenched disorder to evolve along one chosen direction only. This anisotropic disorder is admittedly an approximation, but it is known to reproduce correctly the qualitative features of the Anderson transition in the absence of decoherence, as shown  in the earlier works \cite{Kumar86, Shapiro86}. The probability density $P^{(d)}(\rho,l)$ of the resistance of a d-dimensional hypercubic sample is accordingly found to obey the integro-differential evolution equations
\bea
\f{\pa \chi^{(d)}}{\pa \ln l}&=&-(d-1)x\f{\pa\chi^{(d)}}{\pa x}+[(x^2+Dx)\f{\pa^2\chi^{(d)}}{\pa x^2}+(2x-Dx-x^2)\f{\pa \chi^{(d)}}{\pa x}-x\chi^{(d)}]~ l~, \label{momgend}\\
\f{\pa K^{(d)}}{\pa \ln l}&=&-(d-1)x\f{\pa K^{(d)}}{\pa x}+\left[(x^2+Dx)\f{\pa^2K^{(d)}}{\pa x^2}+(x^2+Dx)(\f{\pa K^{(d)}}{\pa x})^2+(2x-Dx-x^2)\f{\pa K^{(d)}}{\pa x}\right.\nn \\
&-&\left.x\right] ~l~, \label{cumgend}
\eea
where $l$ in the above equations is given by the integral in Eq.(\ref {lrelation}), but with $\rho^{(1)}_1$ in the integrand now re-interpreted as $\rho^{(d)}_1$. Clearly, in the limit $D=0$, the above equations for the generating functions reduce to the corresponding Eqs.(6,7) of Ref.\cite{Kumar86}.

In particular the fixed point probability distribution for $d=3$ obtained by setting $\pa \chi^{(d)}/\pa \ln l=0$ and inverting the Laplace transform of the solution for $\chi^{(d)}$ is nothing but the known fixed point power-law distribution \cite{Shapiro86}. 

In the presence of decoherence $(D\ne0)$, however, there is no fixed point even for arbitrarily small values of $D$ for $d=3$. In order to see this, consider the evolution equation for the first cumulant $K^{(d)}_1 (\equiv \rho^{(d)}_1)$ obtained from the cumulant-generating Eq.(\ref {cumgend})
\bea
\f{\pa K^{(d)}_1}{\pa \ln l}&=&-(d-1)K^{(d)}_1+[1+2K^{(d)}_1-DK^{(d)}_1-D{K^{(d)}_1}^2-DK^{(d)}_2]\nn \\ 
&&\int^{\rho^{(d)}_1}_0\f{d\tilde\rho^{(d)}_1}{-\f{4}{3}D{\tilde\rho^{(d)^3}_1}-2D{\tilde\rho^{(d)^2}_1}+(2-D)\tilde\rho^{(d)}_{1}+1}~,  \label{cum1d}
\eea
where we have replaced the length $l$ in terms of $\rho^{(d)}_1$ as explained above. Carrying out the integration occurring in Eq.(\ref {cum1d}) numerically (using Mathematica), we found no solution with ${\pa K^{(d)}_1}/{\pa \ln l}=0$ for any non-zero value of $D$ however small (down to $D \sim 10^{-6}$) confirming that there is no fixed point. This should, of course, be physically so inasmuch as the decoherence is expected to suppress  quantum interference effects (and localisation), in the limit of large sample size. For $D\ne 0$, however, we do expect the probability density to vary slowly in the vicinity of the $D=0$ fixed point, now become a crossover. Indeed, setting $\pa \chi^{(d)}/\pa \ln l \simeq0$ for small non-zero $D$, we obtain for the quasi-fixed-point probability density of resistance 
\bea
P(\rho^{(d)}_{1})&=&\f{D^{1-\alpha}~e^{-D(1+\rho^{(d)}_{1})}~(1+\rho^{(d)}_{1})^{-\alpha}}{\Gamma(1-\alpha,D)}~,\label{pden}\\
{\rm where}~~~\Gamma(1-\alpha,D)& \equiv &\int^{\infty}_{D}e^{-u}u^{-\alpha} du ~~~{\rm and}~~~\alpha=\f{d-1}{l|_{\rho^{(d)}_{1}}}~.\nn
\eea
Here $\rho^{(d)}_1~(\simeq \rho^{(d)\ast}_1=1.96~{\rm for}~ d=3)$ is the average resistance corresponding to the quasi-fixed-point 
probability density, and $l|_{\rho^{(d)}_1}$ is the value of the integral Eq.(\ref {lrelation}) with the upper limit ${\rho^{(d)}_1}$. It is clear from Eq.(\ref {pden}) that a non-zero value of $D$ (decoherence) makes all the resistance moments finite, that is it cuts-off the otherwise divergent resistance fluctuations. In the absence of decoherence (D=0), Eq.(\ref{pden}) gives a
power -law probability distribution for resistance at the mobility
edge as in Ref.\cite{Shapiro86}. It is to be noted, however, that all numerical
work on tight-binding Anderson model shows that the distribution of
conductance in 3d at the mobility edge (the fixed point) is far from
a power law \cite{Markos99, Soukoulis99}. We think that this could
be for two reasons: First, the neglect of the transverse fluctuations
in our anisotropic Migdal-Kadanoff procedure,
 and secondly as the numerical results are all for the ensemble averaged
two-probe conductance (Tr $tt^{\dagger}$) (where $t$ is the transmission matrix) while we have
calculated the ensemble averaged four-probe resistance. It is to be noted here that while the four-probe resistance is  unbounded from above and can, therefore,  have large fluctuations, the two-probe conductance is by definition bounded from above and can fluctuate relatively much less. Thus, e.g., in the 1d case, Tr $tt^{\dagger} \leq 1$; but, of course, there is no fixed point in the 1d case.  We would like to point out here that the invariant imbedding equation is, of course, known for d-dimensional as also for the quasi-one dimnsional case (see Ref.\cite{Rammal87}, Eq.2.28), but an analytical solution is
lacking,
making any comparison with the available results for the quasi-1d, D=0 conductance distribution \cite{Muttalib99, Garc01} impossible. Our immediate interest, however, lies in the fact that decoherence cuts off the resistance
fluctuations exponentially.

Finally, we consider the asymptotic behaviour of the resistance in 3 dimensions  in the presence of decoherence in the metallic regime as the sample size tends to infinity. In 3 dimensions with $D\ne0$, we expect the resistance to tend to  a small value in the mean  along with a narrow width (the variance) of the distribution. This motivates us to approximate the evolution Eq.(\ref {cum1d}) for the first moment as
\bea
\f{\pa {\rho}^{(d)}_1}{\pa \ln l}=-(d-1){\rho}^{(d)}_1+[1+(2-D){\rho}^{(d)}_1]\int^{\rho^{(d)}_1}_0\f{d\tilde\rho^{(d)}_1}{1+(2-D)\tilde\rho^{(d)}_{1}}~. \label{mom1d}
\eea
 Now, consider first the 3-dimensional case (d=3) in the metallic regime starting with the resistance $\rho_0={\rho}^{(3)}_1(l_0)$ at a length scale $l_0$. Let this evolve through Eq.(\ref {mom1d}) to a length scale $l>>l_0$ with $\rho^{(3)}_1(l)\equiv \rho<<\rho_0$. Eq.(\ref {mom1d}) then gives  
\bea
\int^{\rho^{(3)}_1}_{{\rho}_0}\f{d\tilde\rho^{(3)}_1}{-\tilde{\rho}^{(3)}_1+\f{2-D}{2}{\tilde\rho^{(3)^2}_1}}=\ln(\f{l}{l_0})~, 
\eea
or, in term of the conductivity ${\sigma}^{(3)}(l) \equiv g/l,~ g \equiv 1/\rho ~{\rm and}~ g_0 \equiv 1/{\rho}_0$, we have 
\bea
{\sigma}^{(3)}(l)=\f{g_0 -1}{l_0} + \f{1}{l} +\f{D}{2}(\f{1}{l_0}-\f{1}{l})~.\label{mom2d}
\eea
Equation (\ref {mom2d}) clearly shows that increasing decoherence (D) increases the metallic conductivity in 3 dimensions. Indeed, one can re-write the correction $D/2 l_0$ as $1/L_{\phi}$ with $L_{\phi}$ a phase-cut-off (dephasing) length scale as usual. Proceeding in similar way, we get for the 2-dimensional case a logarithmic correction to the conductivity $\sigma^{(2)}(l)$ (noting that in 2 dimensions conductivity is the same as conductance)
\bea
\sigma^{(2)}(l)=\sigma_0+\f{D-2}{2}\ln(\f{l}{l_0})~,
\eea
where $\sigma_0$ is the conductivity (or the conductance) at the starting length scale $l_0$. Again, the conductivity $\sigma^{(2)}(l)$ is seen to increase with increasing decoherence $D$. This is qualitatively consistent with the negative temperature coefficient of resistance observed in disordered conductors at low temperatures in the weak localisation regime, in particular for $2d$ systems \cite{TVR85}.

\section{Discussion}
\label{sec:disc}
We have extended the phenomenology of decoherence known well in the context of phase-sensative systems, such as mesoscopic rings \cite{Roy08} and 1-dimensional quantum wires \cite{Datta89, Datta91, DibAbhi07, Pastawski90, Maschke91, Maschke94}, to higher dimensions -- specifically to a d-dimensional disordered conductor for $d=2~{\rm and}~3$. Our treatment here follows the invariant embedding approach developed earlier \cite{Kumar85, Heinrichs86, Rammal87}, beginning with the 1d case. It treats decoherence and disorder formally $at~par$ in that the two are introduced through appropriately chosen and parametrised scattering matrices distributed over the conductor. The invariant embedding appraoch is naturally suited to the problem on hand as it gives the evolution-in-length of the resultant emergent quantities such as the reflection coefficient related directly to the Landauer 4-probe resistance of interest. Decoherence is realised specifically through  stochastic absorption of the wave-amplitude into distributed side (transverse) channels, and the subsequent re-injection of the absorbed fraction back into the conductor so as to add incoherently to the (longitudinal) coherent transport. This is essentialy in the spirit of B{\"u}ttiker's reservoir-induced-decoherence. A point to note here is that the current-conserving re-injection is realised here self-consistently  through the use of the 4-probe resistance which now needs to be calculated with the coherent-only reflection coefficient. Extension to higher dimensions has been carried out within the Migdal-Kadanoff procedure assuming the disorder to evolve only along an arbitrarily chosen direction for the current. This choice of anisotropic disorder is admittedly an approximation, but its innocuous nature is borne out $a~ posteriori$ by the fact that this approximation had correctly given the unstable fixed point for the disorder induced Anderson (metal-insulator) transition for $d=3$ in the absence of decoherence. Its reasonableness may be attributed to the transverse mixing up of disorder by the evolution equation. Physically, moreover, the classicalisation expected from decoherence should make the approximation even better. A non-trivial result of our work is the elimination of the unstable (Anderson) fixed point due to decoherence. Again, it is expected on physical grounds that the fixed point should get replaced by a crossover for $D\ne 0$. So is the finiteness of all moments, that is the suppression of resistance fluctuations due to decoherence, as is evident from our Eq.(\ref {pden}). A point to note is the decoherence correction to the quantum conductivity for $d=3$, where a cut-off length (dephasing length) appears naturally.  Finally, we would like to point out here that the decoherence, through stochastic absorption into the transverse channels and the re-injection, does not cause scattering in the coherent longitudinal (transport) channel in the sense of momentum randomisation that would have  given additional resistance. Indeed, as is clear from our  Eq.(\ref {II2}), in the absence of scattering by disorder, the reflection amplitude $(R)$ remains identically zero for all lengths, independently of the value of $\eta$ (that parametrises decoherence). This is also obvious from the Eq.(\ref {SmatD}). We would aptly like to call this a $pure$ decoherence without any concomitant elastic scattering.

% \bibliography{ref}

\end{document}